# Sistem Informasi Geografis Ruang Terbuka Hijau Kawasan Perkotaan (RTHKP) Palembang

Andika[1], Leon Andretti Abdillah[2], Muhammad Ariandi[3]

[1,2,3] Program Studi Sistem Informasi, Fakultas Ilmu Komputer, Universitas Bina Darma
Palembang, Indonesia
[1] andidika459@yahoo.co.id, [2]leon.abdillah@yahoo.com

**Abstract.** Geographic Information System (GIS) is a computer-based system used to store and manipulate geographic information. In this study, GIS is used to obtain information about "open green space of urban areas" (RTHKP). Office of street lighting and Cemetery Palembang is one agency that regulates the green open spaces but not using media such as websites that support the community to get information about the open green space of urban areas in the city of Palembang, so it was apparent from the author will build a system RTKHP with geographic information system development methodology Rational Unified Process (RUP), the programming language PHP and uses a MySQL database. With the GIS open green space of urban areas that will be made later can help facilitate the public to get information related to RTHKP and assist the Department of street lighting and landscaping burial in managing and providing related information RTHKP in Palembang so that delivery of information to be more effective.

## 1  Pendahuluan

Teknologi informasi saat ini telah merubah suatu pola pikir penggunanya untuk menciptakan suatu aplikasi yang awalnya sulit hingga menjadi lebih mudah. Hal ini disebabkan oleh beragamnya teknologi yang digunakan untuk melakukan berbagai aktivitas dan kebutuhan sehari-hari. Disamping itu, TI juga mampu dikolaborasikan dengan banyak bidnag ilmu yang berbeda [1]. Salah satunya adalah penggunaan sistem informasi geografis (SIG) atau biasanya dikenal dengan *geographic information system* (GIS). GIS merupakan suatu sistem berbasis komputer yang digunakan untuk menyimpan dan memanipulasi informasi-informasi geografis. GIS dirancang untuk mengumpulkan, menyimpan dan menganalisis objek-objek dan fenomena-fenomena dimana lokasi geografis merupakan karateristik yang penting dan kritis untuk dianalisis [2].

Dengan berbagai kemudahan yang diberikan dari penggunaan sistem informasi geografis tersebut, Salah satunya adalah kemudahan dalam mendapatkan informasi mengenai ruang terbuka hijau kawasan perkotaan (RTHKP). Menurut Peraturan Menteri Dalam Negeri Nomor 1 Tahun 2007, tentang Penataan RTHKP, bahwa ruang terbuka adalah ruang-ruang dalam kota atau wilayah yang lebih luas baik dalam bentuk area atau kawasan maupun dalam bentuk area memanjang atau jalur dimana dalam penggunaannya lebih bersifat terbuka yang pada dasarnya tanpa bangunan. RTHKP selain sebagai kawasan lindung juga berfungsi sosial yakni sebagai *open public space* untuk tempat berinteraksi sosial dalam masyarakat seperti tempat rekreasi, sarana olahraga dan area bermain. Informasi terkait RTHKP dapat dilakukan dengan menggunakan GIS untuk mendapatkan informasi di wilayah tersebut.

Umumnya, masyarakat Palembang hanya mengetahui beberapa ruang terbuka hijau yang sudah familiar yang ada di tengah Kota Palembang tetapi tidak mengetahui bahwa ada tempat-tempat lain yang menjadi tujuan ruang terbuka hijau favorit. Untuk menutupi kendala masyarakat Palembang dalam mencari ruang terbuka hijau maka dirancanglah sistem informasi geografis yang bisa memberikan informasi tentang ruang terbuka hijau yang ada dikota palembang sehingga masyarakat bisa dengan mudah mengunjungi dan mencari alternatif lain ruang terbuka hijau di Palembang. Dalam penelitian ini penulis akan memetakan lokasi RTHKP berdasarkan peta *layout* Palembang (gambar 1).

Sejumlah penelitian telah penulis kaji untuk pengembangan SIG ini, antara lain: 1) Sistem informasi geografis lokasi perumahan berbasis android [3], dan 2) Analisis Ruang Terbuka Hijau Kota Semarang Dengan Meggunakan SIG [4], dan 3) Aplikasi Sistem Informasai Geografis (SIG) dalam Analisis Pemanfaatan dan Pengelolaan Ruang Terbuka Hijau Kota (RTHK) [5].

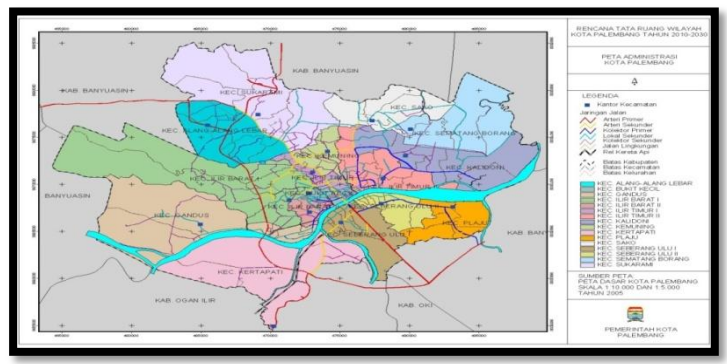

**Gambar 1.** Peta *Layout*

## 2  Metode Penelitian

Penelitian ini dilakukan di Dinas Penerangan Jalan Pertamanan dan Pemakaman Kota Palembang. Metode pengumpulan data yang digunakan dalam penelitian ini adalah: 1) Wawancara dengan pihak dinas terkait, 2) Observasi dengan cara mengamati langsung keadaan dan kegiatan pada objek guna mendapatkan keterangan yang akurat, dan 3) Kepustakaan.

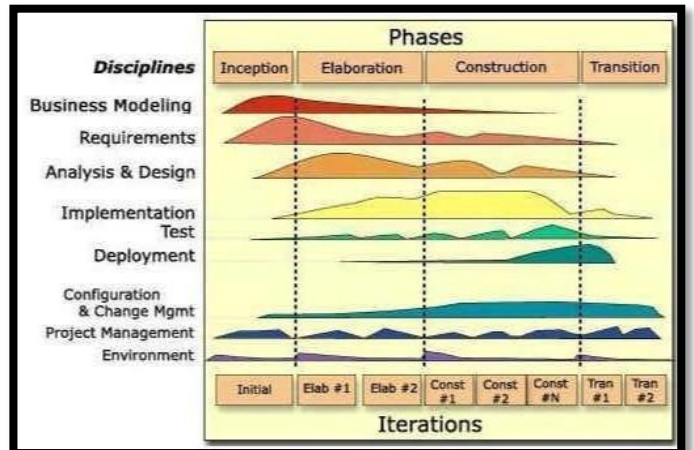

**Gambar 2.** Model Pengembangan Sistem *Rational Unified Process* (RUP)

Metode pengembangan yang digunakan pada penleitian ini adalah *rational unified process* (RUP) [6] adalah perspektif, akurat sistem pembinaan, sering digunakan untuk mengembangkan sistem berdasarkan objek atau komponen dasar teknologi. RUP terdiri atas 4 (empat) fase utama, yaitu: 1) *Inception*, 2) *Elaboration*, 3) *Construction*, dan 4) *Transition* (gambar 2).

## 3  Hasil dan Pembahasan

Berdasarkan hasil penelitian yang telah dilakukan, perancangan dan berakhir dengan pembuatan program yang sesungguhnya, maka hasil yang dicapai oleh penulis adalah sebuah Sistem Informasi Ruang Terbuka Hijau Kawasan Perkotaan (SI-RTHKP) Palembang. Sistem ini terdiri: home, profil, taman kota, taman wisata alam.

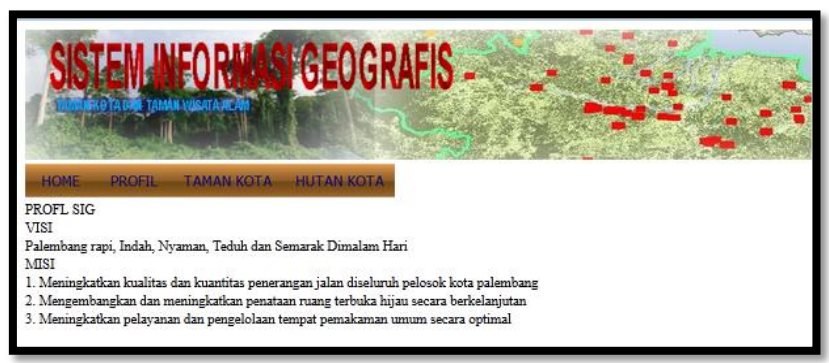

**Gambar 3.** Halaman Profil

### 3.1 Halaman Depan dan Profil

Halaman depan adalah halaman utama dari program dari sistem informasi geografis ruang terbuka hijau. Pada halaman ini terdapat menu home, profil, taman kota, taman wisata alam (gambar 3).

### 3.2 Halaman Peta Taman Kota

Halaman ini berisikan *layout* peta RTHKP Palembang. Pengunjung dapat melihat *layout* peta RTHKP Palembang dengan membuka menu Taman kota (gambar 4). Taman Kota berisikan 10 (sepuluh) Taman Kota, yaitu: 1) Taman Gelora Sriwijaya, 2) Taman Dekranasda, 3) Taman Kampung Kapiten, 4) Taman Benteng Kuto Besak, 5) Taman Monpera, 6) Taman Bawah Jembatan Ampera, 7) Taman Masjid Agung, 8) Taman Kambang Iwak, 9) Taman Masjid Taqwa, dan 10) Taman POM lX.

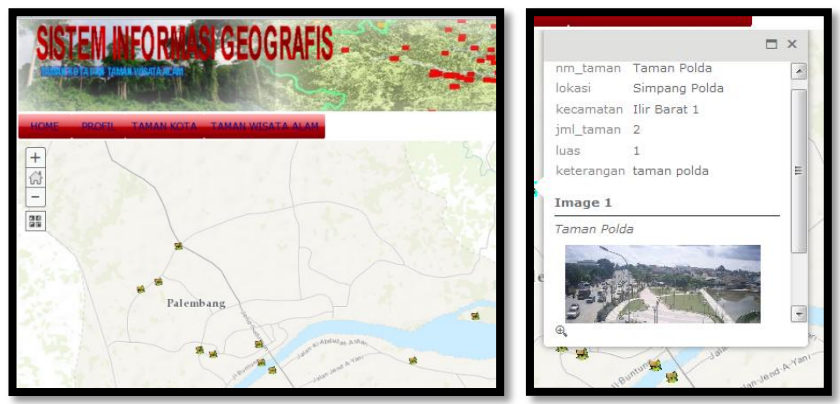

**Gambar 4.** Halaman Peta Taman Kota

### 3.3 Halaman Taman Wisata Alam

Pada halaman taman wisata alam berisikan *layout* peta ruang terbuka hijau kota palembang. Pengunjung dapat melihat *layout* peta ruang terbuka hijau kota palembang dengan membuka menu Taman Wisata Alam. Pada Taman Wisata Alam Berisikan 2 Taman Wisata Alam (gambar 5), yaitu: 1) Taman Wisata Alam Punti Kayu, dan 2) Taman Wisata Alam Pulau Kemaro.

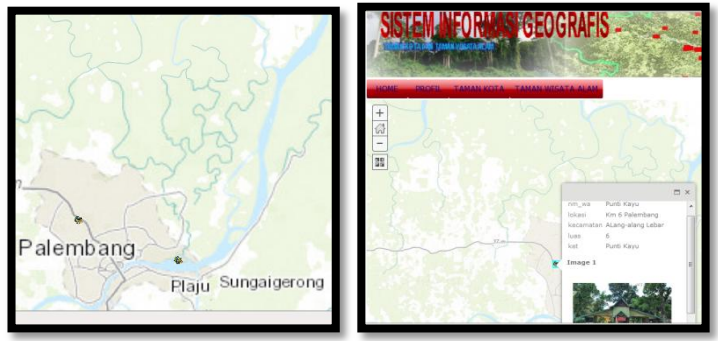

**Gambar 5.** Halaman Peta Taman Wisata Alam

### 3.3 Halaman Admin Kategori *My Content*

Halaman ini berisikan tentang admin setelah login, halaman ini digunakan untuk menginput data peta, menambah lokasi ruang terbuka hijau dan dapat diakses oleh admin.

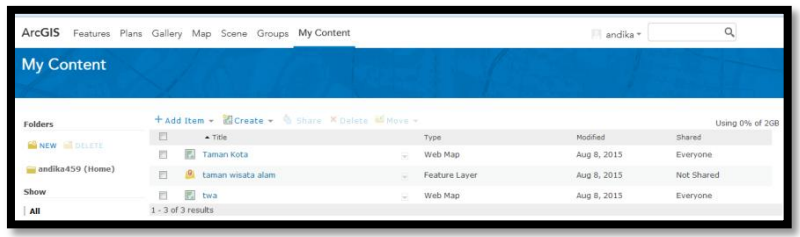

**Gambar 6.** Halaman Admin Kategori *My Content*

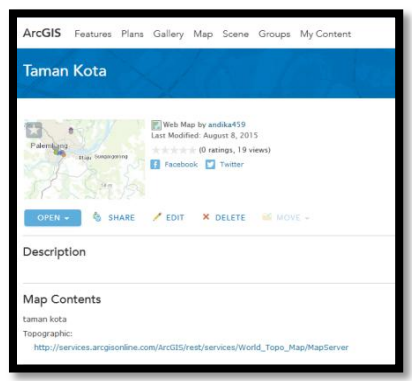

**Gambar 7.** Halaman Admin Data Taman Kota

### 3.3 Halaman Admin Data Taman Kota

Halaman ini berisikan tentang admin data taman kota, halaman ini digunakan untuk menginput data taman kota, menambah lokasi ruang terbuka hijau dan dapat diakses oleh admin.

## 4 Kesimpulan

Berdasarkan hasil penelitian yang penulis lakukan dengan menggunakan pendekatan RUP, maka didapat sejumlah kesimpulan sebagai berikut:
1. SIG yang dihasilkan memberikan informasi sebaran RTHKP di Palembang.
2. Menyajikan data lokasi peta taman kota dan wisata alam.
3. Mempermudah admin memberitahukan informasi RTHKP Palembang.
4. Untuk penelitian selanjutnya, sistem SIG ini bisa memberikan hasil berupa dokumen dengan format standar dokumen global, PDF [7].

## Daftar Pustaka

1. L. A. Abdillah*, et al.*, "Pengaruh kompensasi dan teknologi informasi terhadap kinerja dosen (KIDO) tetap pada Universitas Bina Darma," *Jurnal Ilmiah MATRIK,* vol. 9, pp. 1-20, April 2007.
2. P. Eddy, *Sistem Informasi Geografis: Konsep-Konsep Dasar (Perspektif Geodesi & Geomatika)*. Bandung: Informatika, 2009.
3. L. Novitasari*, et al.*, "Geographic information systems of android-based residential locations," in *4th International Conference on Information Technology and Engineering Application 2015 (ICIBA2015)*, Bina Darma University, Palembang, 2015.
4. H. N. Arifiyanti*, et al.*, "Analisis Ruang Terbuka Hijau Kota Semarang Dengan Meggunakan Sistem Informasi Geografis," *Jurnal Geodesi Undip,* vol. 3, 2014.
5. I. Mildawani and D. Susilowati, "Aplikasi Sistem Informasai Geografis (SIG) dalam Analisis Pemanfaatan dan Pengelolaan Ruang Terbuka Hijau Kota (RTHK) Studi Kasus : Kota Depok," 2012.
6. S. W. Ambler, "A manager's introduction to the Rational Unified Process (RUP)," *Version: December,* vol. 4, p. 2005, 2005.
7. L. A. Abdillah, "PDF articles metadata harvester," *Jurnal Komputer dan Informatika (JKI),* vol. 10, pp. 1-7, April 2012.